\documentclass[11pt]{article}
\usepackage{moriond,epsfig}
\usepackage[latin1]{inputenc}
\usepackage[OT1]{fontenc}

\def\be{\begin{equation}}
\def\ee{\end{equation}}
\def\bea{\begin{eqnarray}}
\def\eea{\end{eqnarray}}

\newcommand{\LCDM}{$ \Lambda $CDM~}

\begin{document}
\title{Measuring dark energy with the integrated Sachs-Wolfe effect}

\author {Tommaso~Giannantonio}

\address {Institute of Cosmology \& Gravitation, University
of Portsmouth,\\ Mercantile House, Hampshire Terrace, Portsmouth, Hampshire, PO1 2EG, UK}

\maketitle\abstracts{I present to this conference our latest measurements of the integrated Sachs-Wolfe (ISW) effect. After a brief review of the reasons for which this effect arises and of the technique to detect it by cross-correlating the cosmic microwave background (CMB) with the large scale structure of the Universe (LSS), I describe the current state of the art measurement. This is obtained from a combined analysis of six different galaxy datasets, and has a significance level of $\sim 4.5 \, \sigma $. I then describe the cosmological implications, which show agreement with a flat \LCDM model with $ \Omega_m = 0.20^{+0.19}_{-0.11}$ at $95\%$ confidence level \cite{Giannantonio:2008zi}. I finally show how these data can be used to constrain modified gravity theories, focusing in particular on the Dvali-Gabadaze-Porrati (DGP) model \cite{Giannantonio:2008qr}.}

\section{Introduction}

The question of dark energy is one of the most interesting open problems in today's physics, and it is important to approach it from different perspectives; at the present, the most studied effects of dark energy are its geometrical consequences on  the cosmic microwave background (CMB), the baryon acoustic oscillation (BAO) and distant supernovae. Another effect of dark energy is the integrated Sachs-Wolfe (ISW) \cite{Sachs:1967er}, which consists of the production of small secondary anisotropies in the CMB, and is caused by the decay of the gravitational potentials which happens if the Universe undergoes a transition from matter domination to a new phase, such as dark energy or curvature (typically at redshift $ z < 2 $).
A direct measurement of these anisotropies is difficult, because their signal is mixed with the primary CMB spectrum which is bigger by an order of magnitude, and also because their amplitude is maximum at the largest angular scales, which are mostly affected by cosmic variance. The measurement of this effect becomes feasible thanks to the technique of cross-correlating the total observed CMB with some tracer of the large scale structure (LSS) of the Universe \cite{Crittenden:1995ak}: the late ISW anisotropies are correlated with the matter density distribution through the effect of the gravitational potential, while the primary CMB can not be so because it has been produced a long time before the formation of structure.

This technique has been used by several groups to detect the ISW effect \cite{ISW} 
 using the WMAP maps of the CMB and several different maps of the LSS out to a median redshift $ \bar z = 1.5 $, obtaining results generally consistent with a concordance \LCDM model and a typical significance of $ 2 - 3 \, \sigma $. To go beyond simple detection and use this method to constrain cosmology, a higher significance is needed, and this can be achieved with a combination of these measurements \cite{combined}.
In this presentation I review the latest results obtained by a full reanalysis of all the relevant datasets \cite{Giannantonio:2008zi}, measuring directly the real space correlation functions and the covariance between them (\S \ref {sec:ana}). Together with a similar analysis in the harmonic space \cite{Ho:2008bz}, this represents the current state of the art for this technique and produces independent constraints on the properties of dark energy or modified gravity theories \cite{Giannantonio:2008qr} (\S \ref {sec:impli}).

\section{Combined analysis} \label {sec:ana}

\begin{figure*}[htp] 
\begin{center}
\includegraphics[angle=0,width=.7\linewidth]{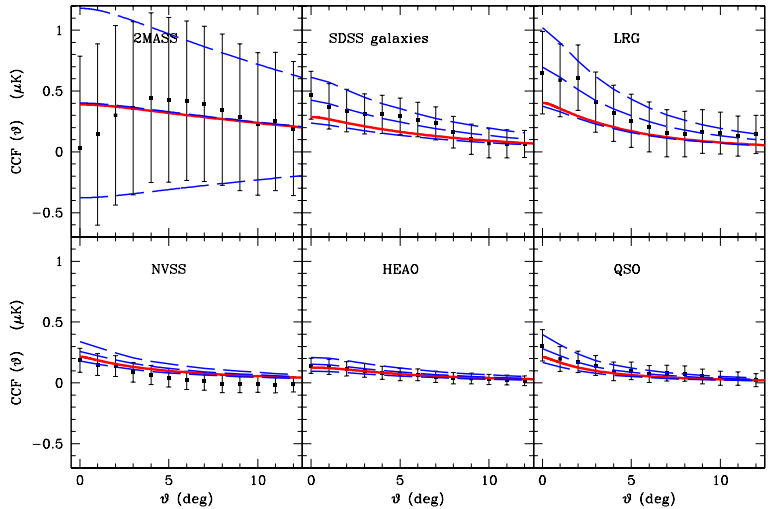}
\caption{Measurements of the cross-correlation functions between all the catalogues and the WMAP CMB maps (black points), compared with the theory from WMAP best fit cosmology and the galactic bias from the literature (red solid lines). The best fit amplitudes and their $1 \, \sigma$ deviations are shown in blue (dashed). }
\label {fig:ccf-all}
\end{center}
\end{figure*}

We perform our analysis correlating the CMB maps from WMAP with six galaxy catalogues: the infrared 2 Micron All-Sky Survey (2MASS), the optical Sloan Digital Sky Survey (SDSS),  the X-ray catalogue from the High Energy Astrophysical Observatory (HEAO) and radio galaxy catalogue from the NRAO VLA Sky Survey (NVSS). Due to the high quality of the SDSS, we can extract three sample from it: the main galaxy sample, the luminous red galaxies (LRG) and a sample of quasars. 
The redshift distribution of these datasets, spanning a range $ 0 < z < 2.5 $, are greatly overlapping, and so are their sky coverages, thus contributing to build a significant covariance between the measurements.
We pixelise all the maps on the celestial sphere using the HEALPix scheme with a resolution of 0.9 deg, and we account for possible foregrounds by masking the most affected areas of the sky, finding that the most severe contamination comes from dust exctinction. We then estimate the two-point cross-correlation functions (CCF) in real space between every pair of maps.  In particular, we look at the auto-correlation functions (ACF) of each catalogue, and we find a good agreement with the galactic biases from the literature. 
We estimate the uncertainties on this measurement in three ways: a Monte Carlo method (MC1), where the covariance matrix is found by measuring the CCF between 5000 random CMB maps and the real observed density maps; a second Monte Carlo method (MC2) which is similar but makes also use of 5000 random density maps, generated including the expected correlations between them and the Poisson noise, and finally a jack-knife error estimation (JK) which is obtained by looking at the variance of the CCF when patches of the data are removed.
The covariances obtained with the three methods are compatible; we choose to use the MC2 method to obtain cosmological constraints since this is our most accurate estimation.
We can see in Fig. \ref{fig:ccf-all} the result of the CCFs with the MC2 errors; 
 we find a good agreement with the theoretical predictions for the \LCDM concordance model, although we note an excess in the signal at the $1 \, \sigma$ level, in general agreement with previous measurements. 
The significance level for each individual catalogue is typically between 2 and $ 3 \, \sigma$, with the exception of 2MASS for which we do not find evidence for a detection. The significance of the total combined measurement is $ 4.5 \, \sigma$ when we include the full covariance matrix between all the data.

\section{Cosmological implications} \label {sec:impli}

If we assume that the observed correlations are produced by the ISW effect, we can obtain cosmological constraints by comparing them with theoretical predictions.
Due to the nature of the CCF, our measurement is largely independent of parameters other than the ones related to dark energy or curvature; we minimise the impact of galactic bias by renormalising it for each model in such a way to impose that the ACF is consistent with the observations.
\subsection {Dark energy}
\begin{figure}[ht] 
\begin{center}
\includegraphics[angle=0,width=0.35\linewidth]{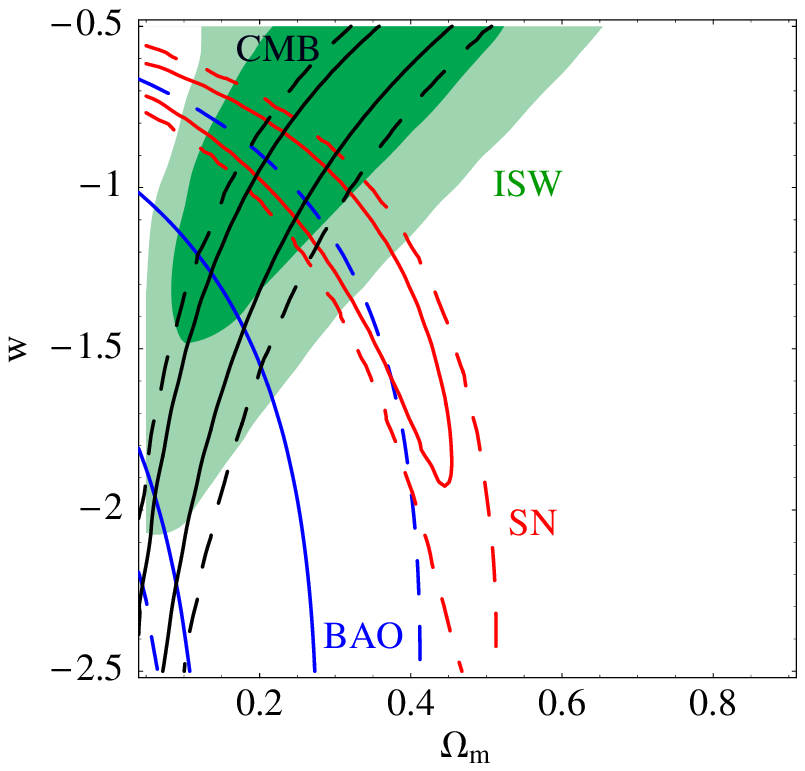} 
\includegraphics[angle=0,width=0.35\linewidth]{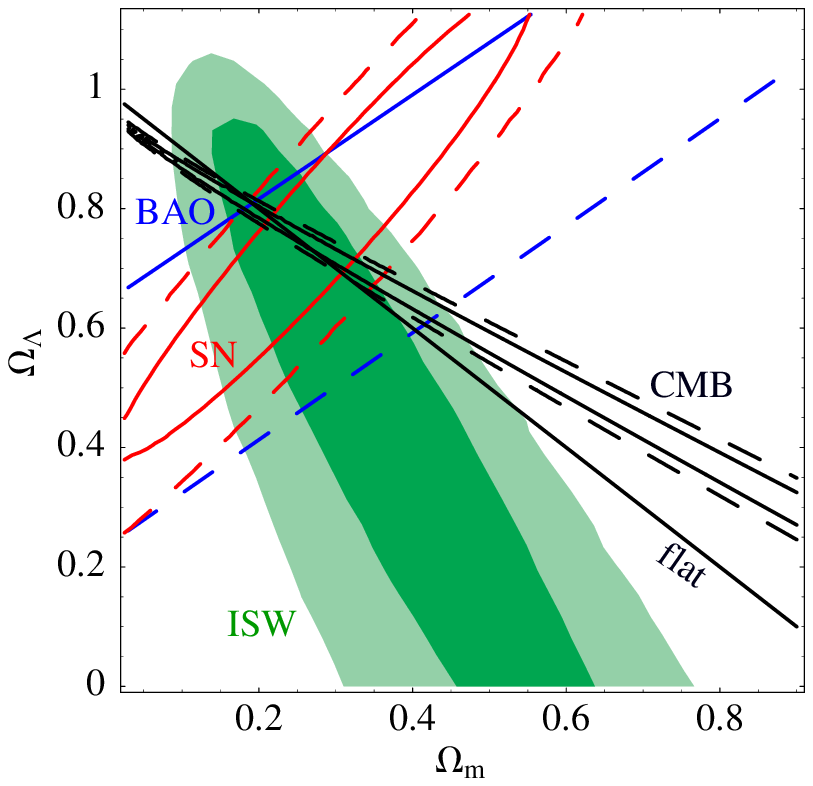}
\caption{Likelihood contours from our ISW data (green) and comparison with constraints from other observations, including CMB shift (black), SNe (red) and BAO (blue) for flat wCDM (left) and curved \LCDM (right) models. 1 and 2 $ \sigma $ contours are shown.}
\label {fig:lik-all}
\end{center}
\end{figure}

We first study a flat \LCDM model with varying $ \Omega_m $: we find the $ 95 \% $ confidence level interval is  $ \Omega_m = 0.20^{+0.19}_{-0.11}$. The low value for this parameter is caused by the excess observed in the CCFs.
 We then extend the analysis to two more classes of models: flat wCDM models with varying both $w$ and $ \Omega_m $, and curved \LCDM models.
We can see the results for these models in Fig. \ref{fig:lik-all}; here we also include the corresponding likelihood contours obtained with other cosmological probes \cite{otherexp}, such as the CMB shift parameter, supernovae Ia and BAO. The result is compatible with the concordance, flat \LCDM model.
In the flat case, we can understand that $ w = -1$ is a good fit by noting that the observed excess in the CCF is largely redshift independent, while models with higher (lower) equation of state would have excess ISW at higher (lower) redshift only. 
In the curved case, it is interesting to notice that the degeneracy direction in the curved case is different from the other experiments; this is due to the fact that also curvature can produce late ISW.

\subsection {Modified gravity}

\begin{figure*}[htb] 
\begin{center}
\includegraphics[angle=0,width=0.35\linewidth]{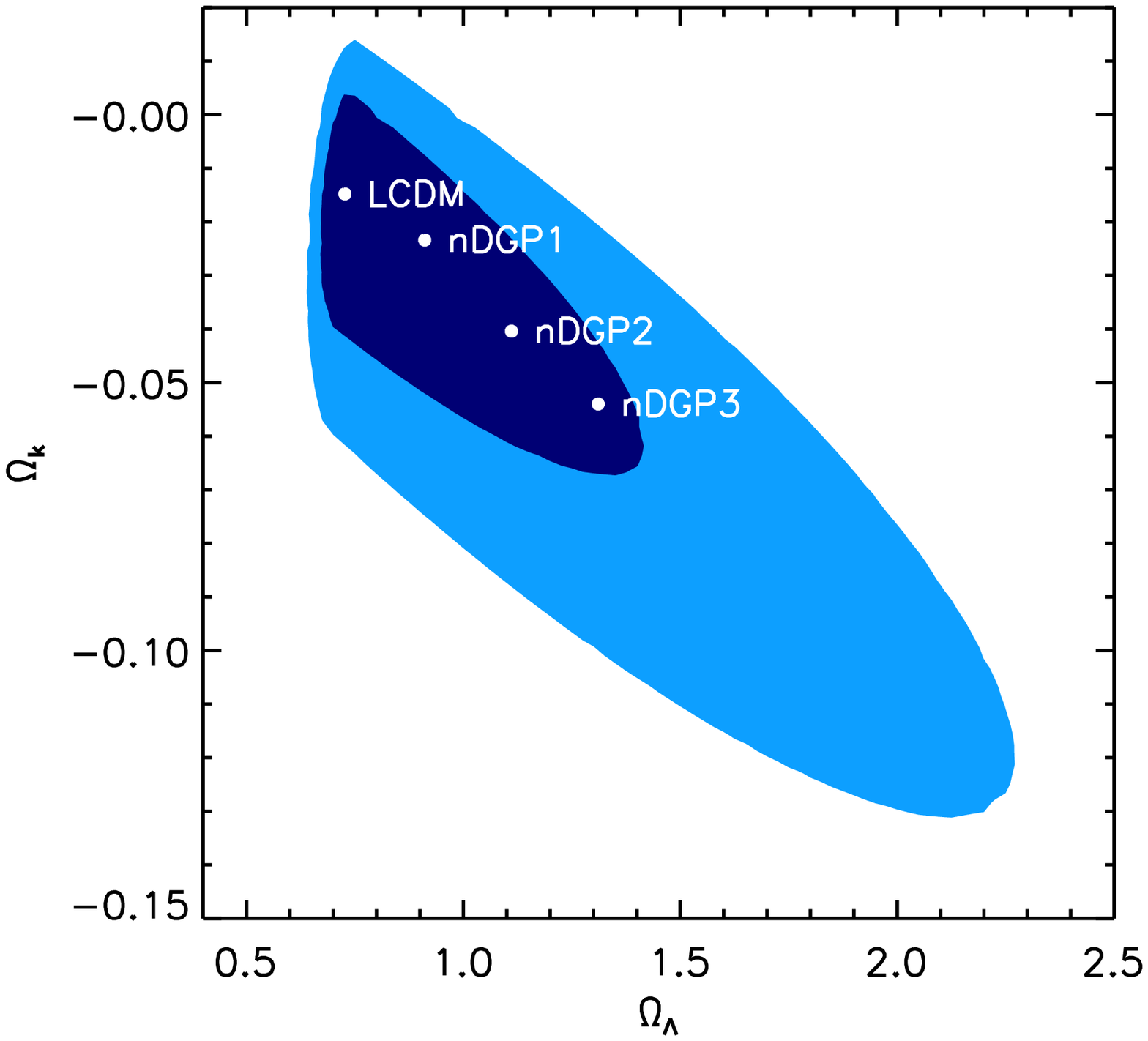} 
\includegraphics[angle=0,width=.64\linewidth]{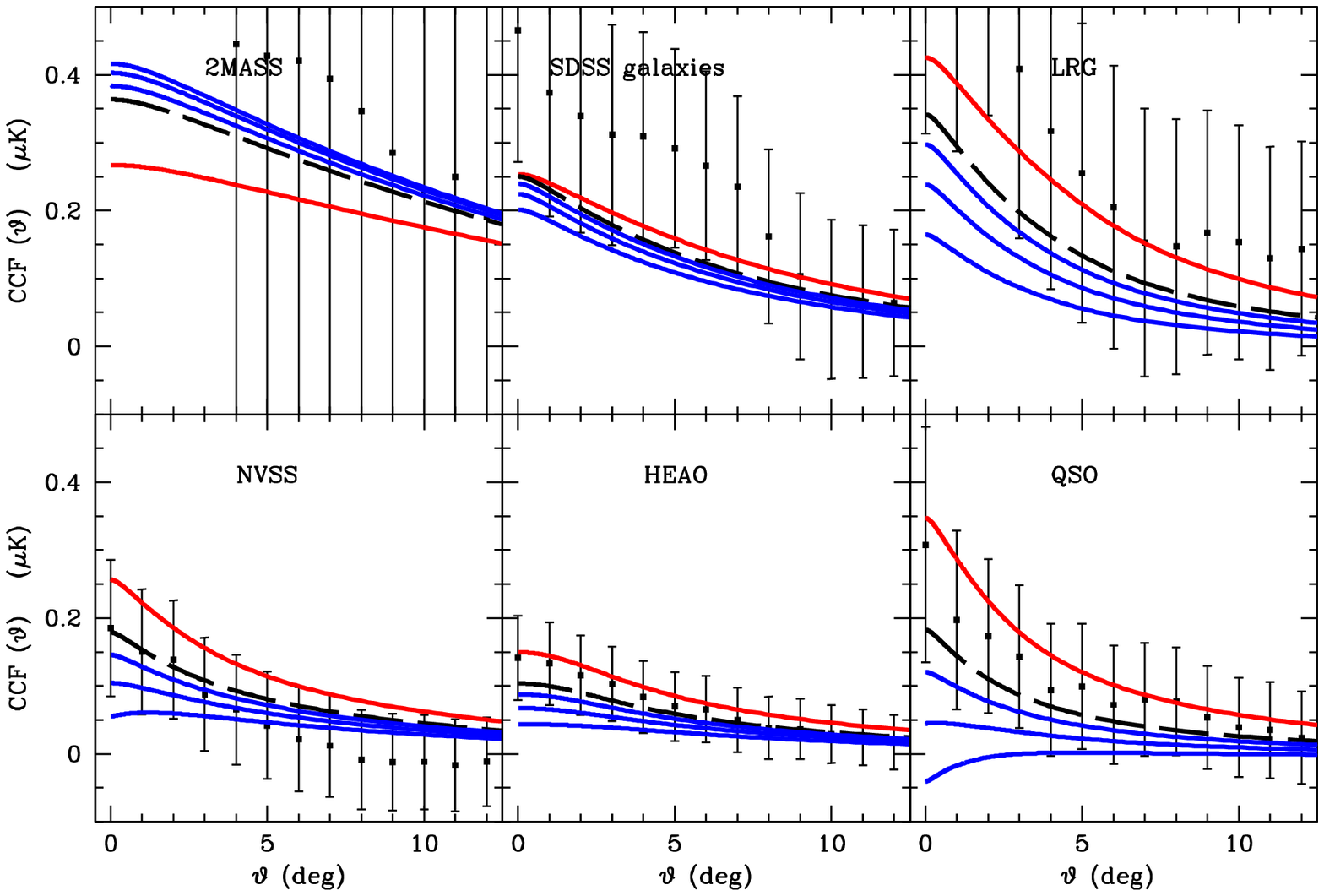}
\caption{Left: likelihood contours for nDGP models from geometrical tests (SN+CMB+$H_0$). Right: Cross-correlation functions for the same four models (nDGP in blue, \LCDM in black) and for the corresponding SA model (red). }
\label {fig:lik-dgp}
\end{center}
\end{figure*}

It is also possible to use the ISW data to constrain modified gravity theories, since in fact, ours is a measurement of the evolution of the gravitational potentials at late times. We focus in particular on the DGP model of gravity \cite{Dvali:2000hr}, which consists of two branches, one of which can explain acceleration without dark energy (self accelerating, SA) while the other still needs a tension on the brane, acting as a cosmological constant (normal branch, nDGP).
From geometrical tests at the background level (SNe, Hubble constant and CMB shift), the SA is in strong tension with the data, while the nDGP is still allowed as we can see from the left panel of Fig. \ref {fig:lik-dgp}.
For this reason, it is interesting to proceed to test the growth of structure at the perturbation level, such as with the ISW. Using the quasi-static approach to the DGP perturbations, we find that the predicted ISW correlations for our datasets are as shown in the right panel of Fig. \ref {fig:lik-dgp}: the SA theory would predict higher correlations, which appear favoured by the data, while the nDGP has lower and negative CCFs, in clear contradiction with our measurement. This shows that the ISW data restrict the range of allowed nDGP models: observing that all three nDGP models in the left panel of Fig.~\ref{fig:lik-dgp} are inside the $1 \, \sigma$ region from the geometry test, we can qualitatively see that the ISW test will produce stricter constraints by noticing e.g. that the quasar CCF alone has a significance level of $ > 2 \sigma $, which means that at least two of the nDGP models will be excluded at above this level.

\section{Conclusions}  \label {sec:concl}
In this presentation I have described the latest measurement of the ISW effect and why this is a useful tool to study dark energy. We have performed a full consistent analysis of six galaxy datasets, finding a detection at the overall significance level of $ 4.5 \, \sigma $. The result is consistent with the concordance \LCDM model, although with an excess of $ \simeq 1 \, \sigma$. I have also summarised how we can apply thee data to constrain different dark energy or modified gravity models; in particular, I showed how nDGP models which are still allowed from background tests are instead ruled out by the ISW data.

\end{document}